# First demonstration of C + L band CDC-ROADM with simple node configuration using multiband switching devices

SHUTO YAMAMOTO[1], HIROKI TANIGUCHI[1], YOSHIAKI KISAKA[1], STEFANO CAMATEL[2], YIRAN MA[2], DAISUKE OGAWA[3], KOICHI HADAMA[3], MITSUNORI FUKUTOKU[3], TAKASHI GOH[4], AND KENYA SUZUKI[4*]

[1]*NTT Network Innovation Laboratories, NTT Corporation, 1-1 Hikarino-oka, Yokosuka, Kanagawa 239-0847, Japan.*
[2]*Finisar Australia, 21 Rosebery Ave, Rosebery NSW 2018, Australia.*
[3]*NTT Electronics Corporation, 1-1-32 Shin Urashima Cho, Yokohama, Kanagawa 221-0031, Japan.*
[4]*NTT Device Innovation Center, NTT Corporation, 3-1 Morinosato Wakamiya, Kanagawa 243-0198, Japan.*

**kenya.suzuki.mt@hco.ntt.co.jp*

**Abstract:** While ultrahigh-baud-rate optical signals are effective for extending the transmission distance of large capacity signals, they also reduce the number of wavelengths that can be arranged in a band because of their wider bandwidth. This reduces the flexibility of optical path configuration in reconfigurable optical add/drop multiplexing (ROADM) networks. In colorless, directionless and contentionless (CDC)-ROADM in particular, the effect reduces the add/drop ratio at a node. Multiband ROADM systems are an effective countermeasure for overcoming this issue, but they make the node configuration more complicated and its operation more difficult. In this paper, we analyze the challenges of C + L band CDC-ROADM and show that optical switch devices that operate over multiple bands are effective in meeting them. For this purpose, we built a C + L band CDC-ROADM node based on C + L band wavelength selective switches (WSSs) and multicast switches (MCSs) and confirmed its effectiveness experimentally. In particular, to simplify the node configuration, we propose a reduction in the number of optical amplifiers used for node loss compensation and experimentally verify its feasibility.

## 1. Introduction

With the recent progress in 5G mobile, IoT, and artificial intelligence, there is a growing demand for optical networks with not only higher capacity but also more flexibility [1-4]. In situations where people move around, such as at sporting events and concerts, and in situations where the amount of information that needs to be processed varies from place to place depending on the time of day, such as Mobility as a Service (MaaS) or computational complexity leveling in high-performance computing, the communication volume varies over time from place to place, creating hot and cold spots in the network. In such situations, it is desirable to dynamically change the communication capacity between each node in the network according to the communication volume. Reconfigurable optical add/drop multiplexing (ROADM) networks operated with software-defined networking (SDN) are attractive solutions for efficient data communication because they provide efficient communication resource usage. At the same time, these applications require optical networks with higher capacity. The distance between data centers on different campuses can be as short as several tens of kilometers and as long as several thousands of kilometers, requiring the transmission of high-capacity data over long distances. High-baud-rate optical signals are effective in increasing capacity over long distances because they contribute to reducing the level of multiplicity of digital coherent signals and are effective in transmitting 400- and 800-Gbps signals over longer distances. In response to the demand for higher baud rates, transmission devices [5] and transmission experiments [6] have been reported in the 100-Gbaud class. However, it seems that the effect of high-baud-rate signals on optical networks has not been discussed yet. The effects of high-baud-rate signals on optical networks can be summarized as follows: high-baud-rate signals occupy a large amount of signal bandwidth, which reduces the number of wavelengths available in ROADM systems. In this paper, we discuss the effects of high-baud-rate signals on optical networks. We also propose a countermeasure to the issue of the reduced number of wavelengths due to high-baud-rate signals by using a multiband ROADM system with switching components supporting both the C and L bands. We also explain the challenges of multiband ROADM networks and experimentally prove that multiband switch devices are effective in meeting them.

## 2. Multiband CDC-ROADM

### *2.1 Impact of high-baud-rate signal on ROADM system*

As mentioned in the introduction, high-baud-rate signal such as 130 Gbaud is expected to extend the transmission distance of high-bit-rate signal, including 800 Gbps. When a high-baud-rate signal is deployed, compared with a low-baud-rate signal at the same transmission rate, the multilevel degree of the signal can be lowered, which is advantageous for lengthening a large capacity channel. However, a disadvantage arises, namely the signal band per channel becomes wider. Figure 1 summarizes the relationship between the signal baud rate and the number of wavelength division multiplexing (WDM) channels that can be located in a band. For example, signals with bit rates of 200, 400, and 800 Gbps have baud rates of 32, 64, and 130 Gbaud for dual polarization 16 QAM, respectively, and the bandwidth occupied is two times larger for 400 Gbps and four times larger for 130 Gbps than for 200 Gbps. The wavelength spacings of the WDM of these signals are expected to be set at 50 GHz for 200-Gbps signals, at 75 [7] or 87.5 GHz [8] for 400-Gbps signals, and at higher than 150 GHz for 800 Gbps. As a result, the channel count per band will be halved or quartered for 400- or 800-Gbps signals. This phenomenon leads to a decrease in the number of channels in WDM, and in particular, in the ROADM system; it reduces the degree of freedom of path connections between the nodes. A countermeasure to this issue is to utilize multiple bands for the networks. The use of the L band in addition to the C band roughly doubles the transmission wavelength range and thus relaxes the WDM channel count problem. This is especially effective for a ROADM system where the number of signal channels is a key performance indicator to make sufficient paths between different nodes.

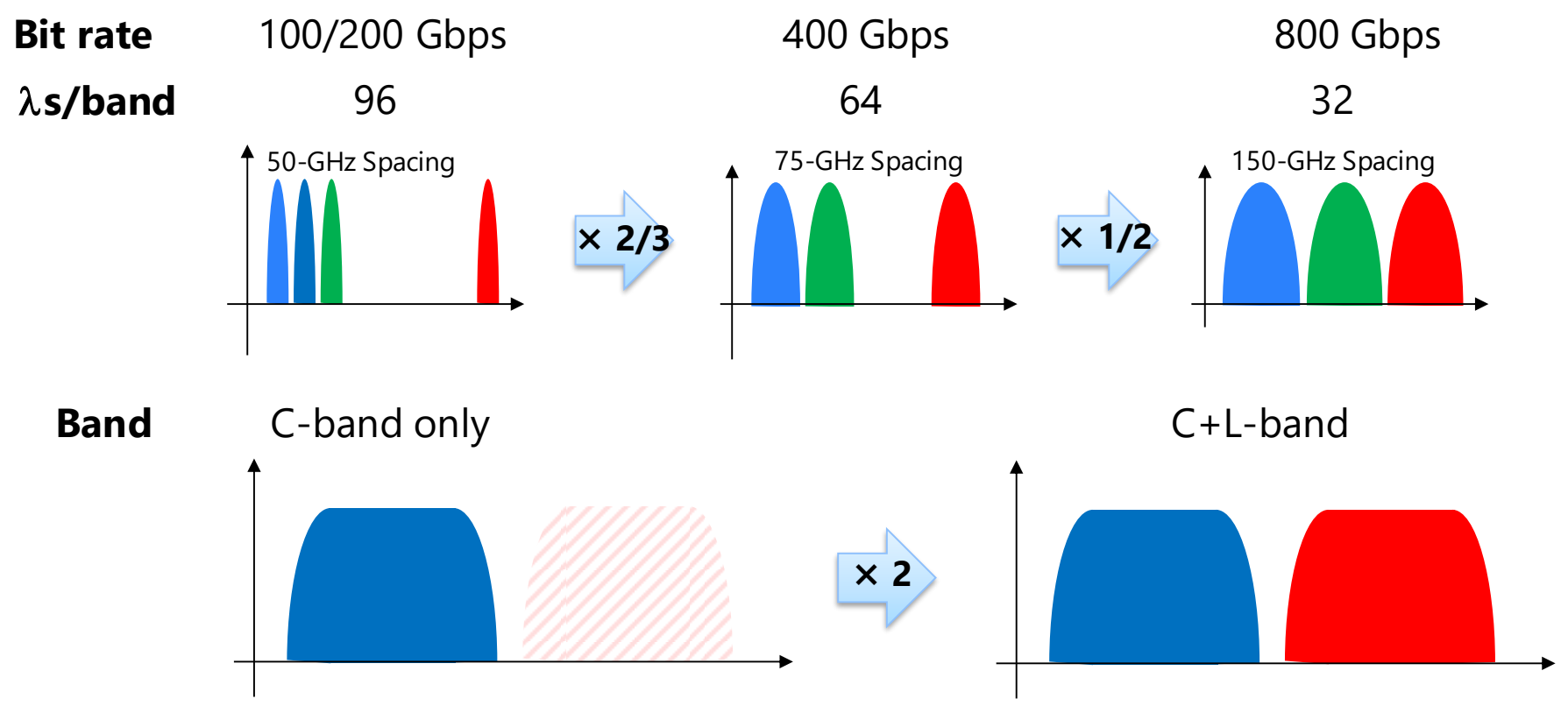

Fig. 1. Impact of higher baud rate signal on the number of WDM channels.

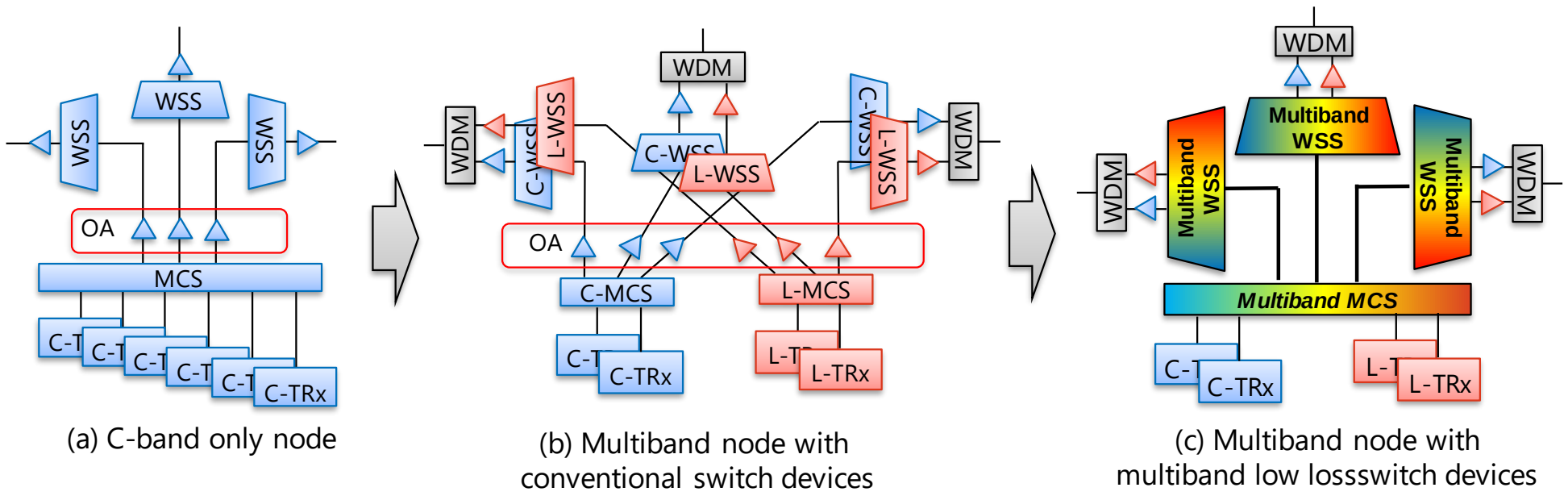

Fig. 2. CDC-ROADM node configuration for (a) C band only, (b) C + L band with conventional switch devices, and (c) C + L band with multiband switch devices.

However, adopting a multiband system raises other problems related to deployment and operation. Currently, the devices for a ROADM system—erbium-doped fiber amplifiers (EDFAs), wavelength selective switches (WSS), and a multicast switch (MCS) for colorless, directionless and contentionless (CDC)-ROADM—are optimized only for a single band. Thus, if we construct a node by using conventional devices, we have to combine separate switch devices as well as separate EDFAs. This increases the complexity of node construction drastically as shown in Fig. 2(a) and (b). The other issue related to operation is the increase of complexity. A ROADM node constructed by separate band devices brings about ambiguity in maintenance. Conventional transponders are also optimized for each band and are only able to transmit either C band or L band signals. When an operator installs additional transponders to the system, they have to make sure they plug the correct transponder into the correct port. Mistakes may interfere with plans to expand transmission capacity.

Multiband switches operating in the C + L band can solve these problems. The node configuration is simplified because switches, which were conventionally separate for the C band and L band, are replaced by integrated devices as shown in Fig. 2(c). From an operational perspective, the MCS functions not only as a transponder aggregator but also as a *band aggregator*, reducing the likelihood of maintenance workers mistakenly inserting or removing transponders.

### *2.2 Effect of multiband operation of CDC-ROADM*

To assess the feasibility of C + L band CDC-ROADMs, we first estimated the availability of the add/drop ratio. To simplify the configuration of C + L band nodes, we should consider optical amplifiers as well as switch devices with multiband operation. Currently, optical amplifiers are optimized for either the C band or the L band, and none can operate in both bands. Therefore, for simplification, we considered reducing the number of optical amplifiers as much as possible. Since optical amplifiers have to be installed to compensate for link loss between nodes, we use two EDFAs, one for the C and the other for the L band, with a configuration sandwiched by two WDM couplers. On the other hand, an amplifier that compensates for the intra-node loss might be unnecessary if the node loss is small enough. In other words, it is essential to reduce the loss of the switching elements. The following two loss factors must be considered to reduce the node loss: the bifurcation loss of the MCS, which is intrinsic in the MCS configuration, and the excess loss of the WSSs and MCS. The former can be lowered by reducing the number of branches in the MCS from the 16 currently used in conventional CDC-ROADM. The latter can be reduced by improving the optical circuit design and manufacturing process.

The challenge in reducing the number of MCS branches is that the achievable add/drop ratio also decreases. However, in the high-baud-rate system, the number of wavelengths that can be included in a band is also reduced. Therefore, a decrease in the add/drop ratio due to a decrease in the number of branches of the MCS has little impact on the ROADM add/drop ratio. In the following, we consider the add/drop ratio by comparing a conventional C band only system with a C + L band system. Table 1 summarizes the add/drop ratio for the ROADM configuration of the C band only system with the conventional 32 Gbaud signal and that of the C + L band one with the 130-Gbaud signal. In the former, it was assumed that the channel spacing is 50 GHz and 96 signals are allocated in the C band; in the latter, the spacing is 150 GHz and 64 signals are used in both the C and L bands. The add/drop ratio also depends on the scale of the WSS, but in this paper, we assume a 1 × 20 WSS in the conventional C band only system, which is available when the conventional system is deployed, while we assume a 1 × 32 WSS in the C + L band ROADM, which is coming into practical use recently. As can be seen from the table, while the add/drop ratio is about 27% in a C band only system with an 8-degree port and 16-client port MCS, which is a standard configuration of conventional CDC-ROADM, the add/drop ratio of 26-% can be obtained even in a C + L band ROADM using a 130-Gbaud signal with a 8-client port MCS. Thus, if the number of branches of the MCS is halved to eight, it can be said that the same operability as the conventional system can be secured. In practice, it is reasonable to estimate the required average add/drop ratio as the total number of wavelengths divided by the number of nodes installed in the ROADM system. Therefore, if the network has ten nodes, the average add/drop ratio is only 10%. It is clear that the cases shown in Table 1 satisfy the condition with the 8-branch MCS.

Table 1. Dependence of add/drop ratio on signal baud rate, node degree. and number of MCS splitting ports.

| | Number of MCS client ports | | | | |
|---|---|---|---|---|---|
| | 4 | 8 | 12 | 16 | 24 |
| C band only system with 32-Gbaud signal | 6.8% | 13.5% | 20.3% | **27.1%** | 40.6% |
| C + L band system with 130-Gbaud signal | 13.0% | **26.0%** | 39.1% | 52.1% | 78.1% |

## 3. Experiment

In this section, we demonstrate the feasibility of a C + L band CDC-ROADM based on two key devices, the C + L band WSS and MCS.

### *3.1 C + L band switches for multiband CDC-ROADM*

The experimental results presented in this paper are based on an early prototype of a liquid crystal on silicon (LCoS) C + L band 1 x 9 WSS. Although both C and L band WSS modules have already been deployed for many years, a WSS module covering both the C and L bands simultaneously has been demonstrated only recently [9]. This represents a major breakthrough that not only supports C + L band installations from day one but also a path to expanding initial C band deployments to the L band when capacity demands. Moving from separate C band and separate L band WSS modules to a fully integrated device that supports both bands simultaneously enables significant power, cost, and space reductions for the next generation of C + L-ROADM systems. In the future, we are planning to scale the design up from a single 1 x 9 WSS to a twin 2 x 32 WSS configuration, which supports the option of having two separate common ports, one for the C band and one for the L band. This will allow the design of compact C + L band route-and-select ROADMs and will be compatible with separate C and L band optical amplifiers, where the C band optical amplifier is connected to the WSS C band common port and the L band optical amplifier is connected to the WSS L band common port.

Figure 3 shows the measured insertion loss distribution of the C + L 1 x 9 WSS used in the experiment. As shown in Fig. 3, the insertion loss ranges from 5.1 to 6.7 dB with an average of 5.5 to 6.1 dB, which is comparable to conventional C or L band only WSSs.

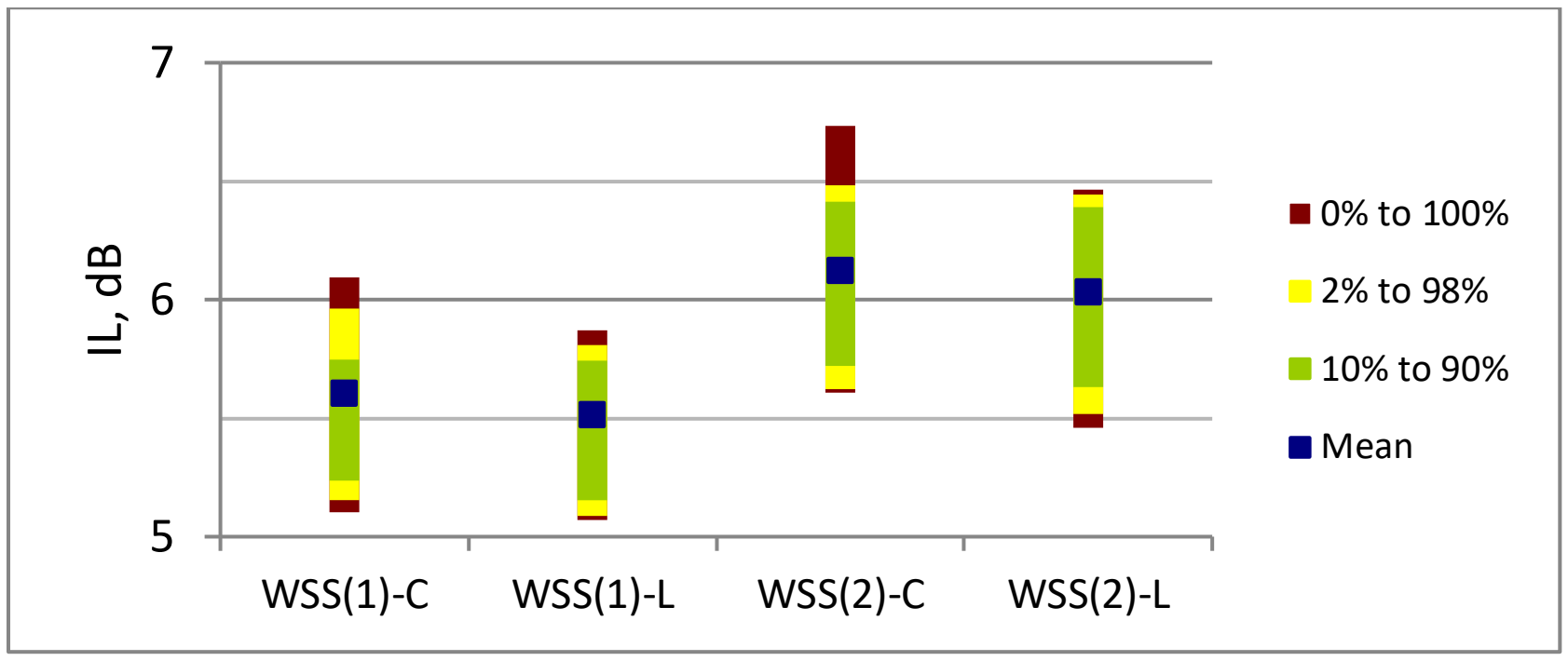

Fig. 3. Measured C + L WSS insertion loss distribution for C and L bands.

As regards the C + L band MCS, we fabricated a 16×8 MCS by using silica-based planar lightwave circuit (PLC) technology [10]. The MCS can be configured in two ways: one is based on spatial optics with microelectromechanical system (MEMS) mirrors as the switching engine, and the other is based on a Mach-Zehnder interferometer (MZI) on a silica-based PLC as the switching engine. In the former, the reflectivity of the MEMS mirrors is determined by the reflectivity of the metal or dielectric mirror, so it can essentially operate over a wide wavelength range. However, because it has mechanical moving parts, a MEMS mirror vibrates when transponders are inserted or removed during equipment maintenance, resulting in the fluctuation of signal power. In addition, it is composed of a free-space optics, which worsens manufacturability. On the other hand, multicast switches using planar waveguides are highly mass-producible because they are manufactured by a process similar to that of CMOS LSIs that uses photolithography, dry etching, etc. However, they suffer from a narrow extinction wavelength range due to the transmission spectrum response of the MZI). We proposed a switch design with good extinction characteristics in a wide wavelength range over the C and L bands. The design utilizes the control of the extinction wavelength of MZIs that constitute an elemental switch [11]. In this work, we developed a 16-degree and 8-client MCS capable of multiband operation with a broadband MZI by using silica-based PLC technology with a 2% refractive index contrast. Figure 4(a) shows the spectrum of the extinction characteristics of the fabricated MCS. The plots show the cumulative port isolation, which is an essential factor for the same wavelength crosstalk in ROADM systems. The cumulative port isolation is the sum of rejection of signals from unexpected client ports leaking into a particular degree port when all client ports are input with the same wavelength, where each signal is configured to be output to a different degree port. As shown in the plot, good characteristics almost over 45 dB were obtained over the C and L bands. The loss of the fabricated MCS, shown in Fig. 4(b), was 11.5 dB on average. The intrinsic loss of the 16×8 MCS is 9 dB. Therefore, the excess loss was only 2.5 dB, indicating that good loss characteristics were obtained.

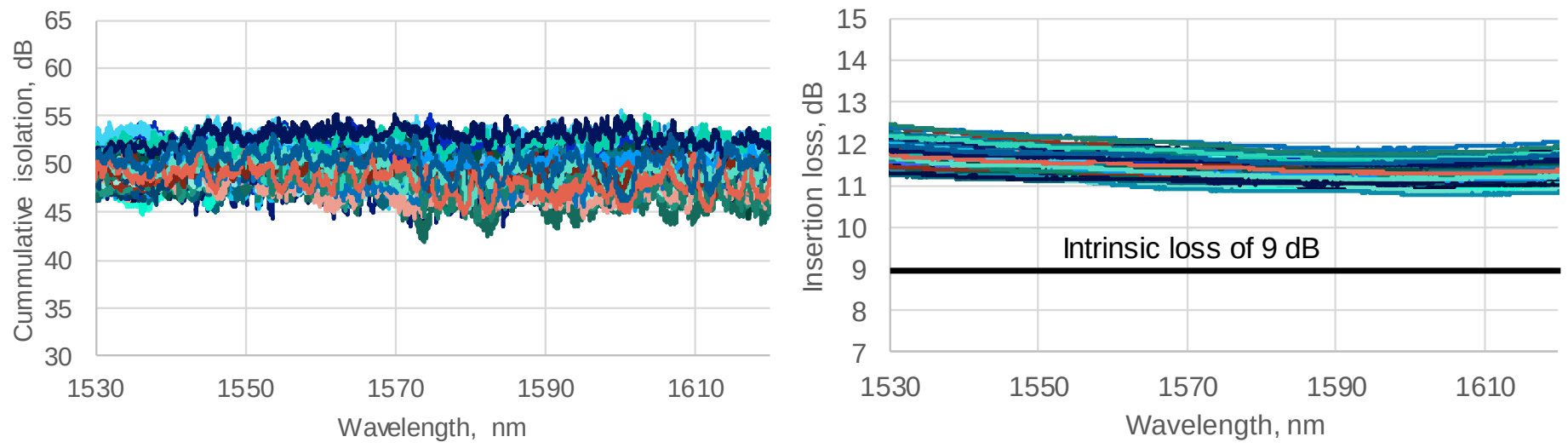

(a) Cumulative isolation

(b) Insertion loss

Fig. 4. Measured spectra of 16×8 MCS. (a) Cumulative port isolation and (b) insertion loss.

### 3.2 Transmission experiments

We conducted a wavelength-routing demonstration for CDC-ROADM in the C + L band for which we used the WSSs and MCSs described in the previous section. Figure 5 shows the experimental setup of our C + L band CDC-ROADM. The configuration has two optical routes. The upper and lower routes correspond to C + L band and C band only links, respectively. The upper link has two of the C + L band WSSs described in the previous section. C and L band optical signals are filtered by the WSSs, whose channel spacing was set at 150 GHz. The lower link has two C band WSSs. The C band signals are filtered by the C band WSSs with a filtering bandwidth of 200 GHz. The reason we use a wider pass band for the C band link is to clarify the impact of the C + L band WSS on the performance. We also installed C + L band MCSs for transponder aggregation, and they select either of the C + L band or C band only links. Two C band real-time coherent transceivers and two L band real-time coherent transceivers generate dual-carrier 1-Tbps super-channels in the C and L bands, respectively. Each dual-carrier 1-Tbps super-channel logically consists of two 500-Gbps signals. The modulation format of the net rate 500-Gbps signal is Nyquist-pulse-shaped 66-Gbaud polarization-division-multiplexed 32 quadrature-amplitude-modulation (PDM-32QAM). The wavelength spacing of the subchannels is 75 GHz, which is the spacing adopted in OIF 400ZR [7].

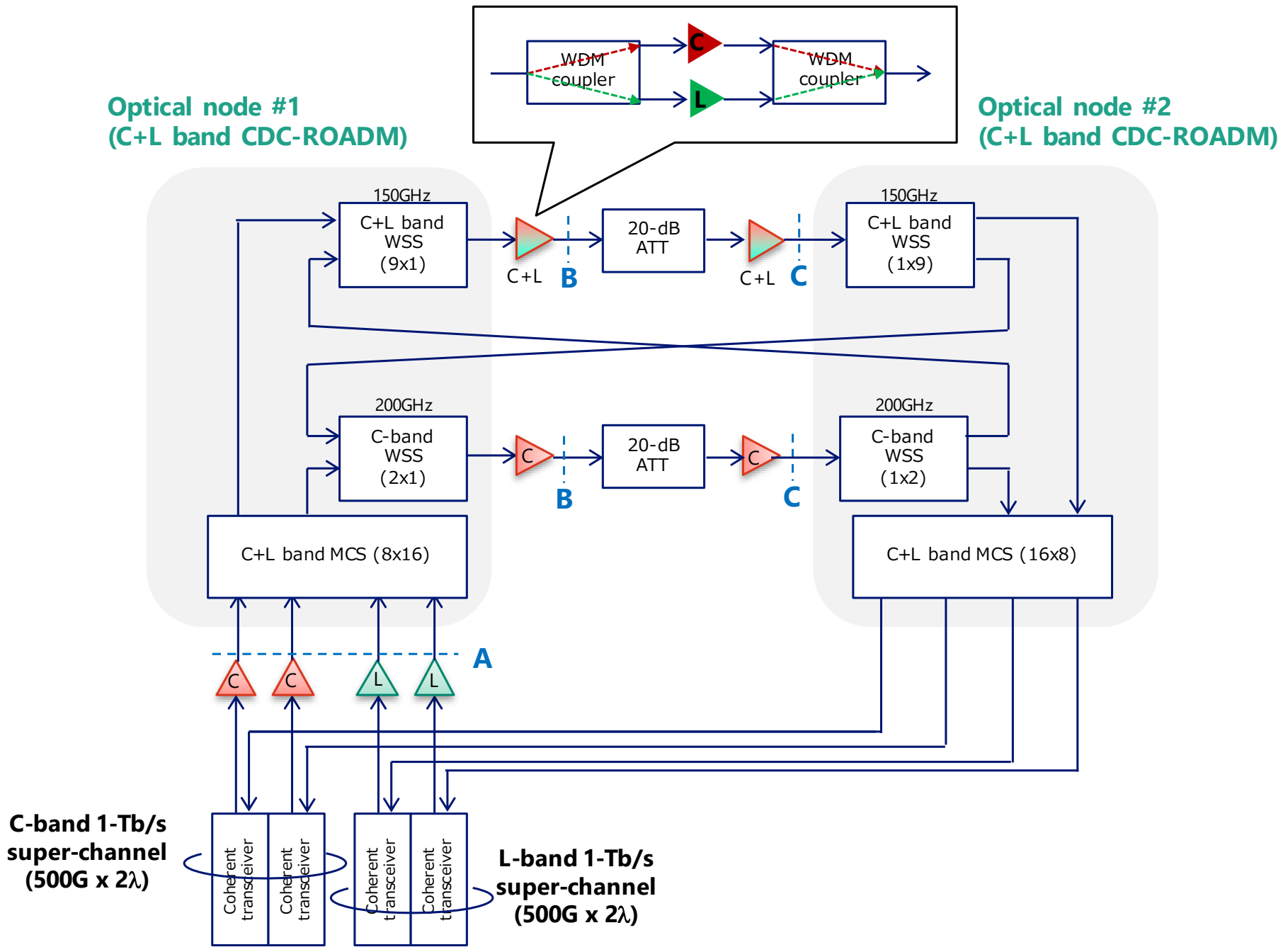

Fig. 5. Experimental configuration using C + L band CDC-ROADM.

To verify the operation of wavelength routing in the C + L band, we transmitted two 1-Tbps super-channels in the C and L bands simultaneously. Figure 6 shows the optical spectra of test signals. There are three cases for the wavelength allocation, and each case has two 1-Tbps super-channels in the C and L bands at the wavelengths near the center or near the edges of the band. The wavelengths are shown in Table 2. The optical signals from the two C band coherent transceivers and the two L band coherent transceivers were input to four of the eight client ports of the 16 × 8 C + L MCS. One degree port of the 16 × 8 C + L MCS was connected to an add-side C + L band WSS, and another degree port was connected to an add-side C band only WSS. The common port of the C + L band WSS was connected to the common port of the other C + L band WSS through two C + L band amplifiers (C band and L band EDFAs were sandwiched between two WDM couplers) and through a 20-dB attenuator emulating the transmission fiber. The same configuration was also prepared for the C band link except that C band EDFAs were installed. In addition, we connected a service port of the C + L band WSS on the add side to a service port of the C band WSS on the drop side and connected a service port of the C + L band WSS on the drop side to a service port of the C band WSS on the add side. These connections allow us to simulate two spans of transmission for the C band signals, one through the C + L band link and the other through the C band only link. It should also be noted here that no amplifier was inserted between the MCSs and WSSs. As mentioned above, as few devices as possible are preferable to simplify the configuration of the CDC-ROADM node. In the 130-Gbaud era, a small amplifier will most likely be integrated in transceivers. Therefore, we place EDFAs between the add-side MCS and transceivers. In the experiment, we fixed the optical power at points A, B, and C at 5, 2, and 5 dBm/subchannel, respectively.

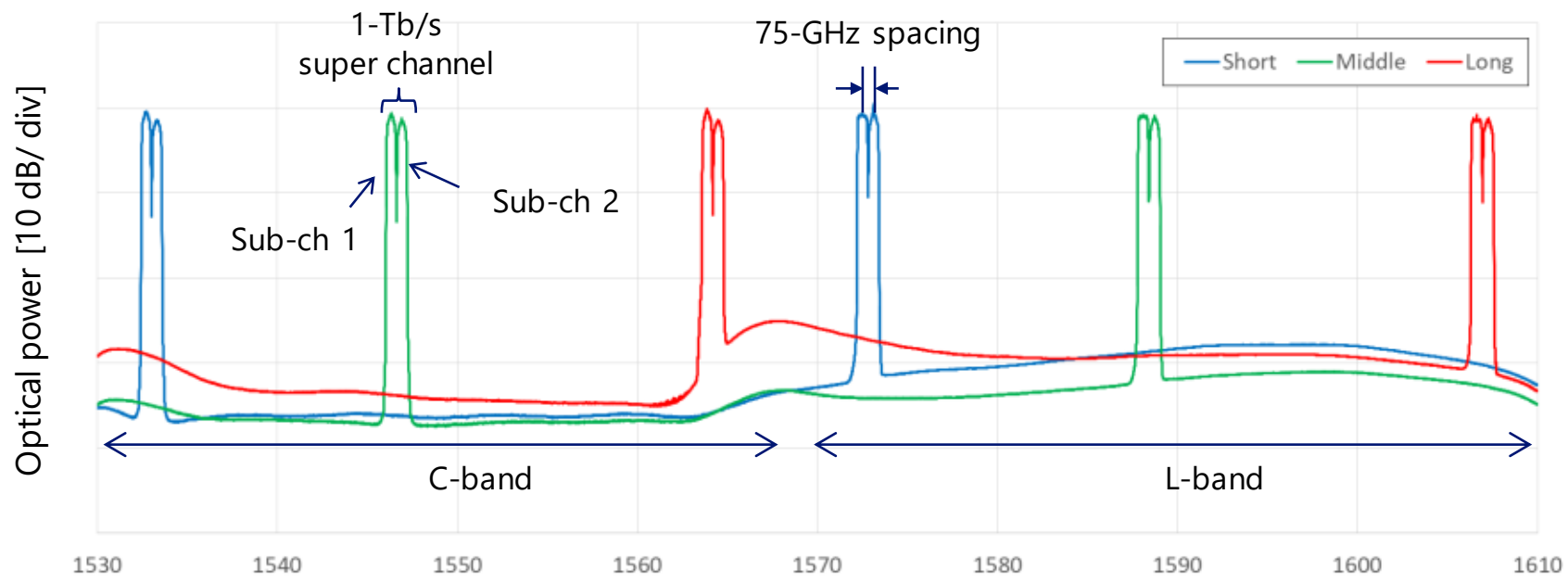

Fig. 6. Test signal spectra. Each spectrum is composed of two subcarriers with 500-Gbps signal.

Table 2. Wavelengths of test signals.

| | C band | | L band | |
|---|---|---|---|---|
| | Subchannel 1, nm | Subchannel 2, nm | Subchannel 1, nm | Subchannel 2, nm |
| Short | 1532.68 | 1533.27 | 1572.48 | 1573.09 |
| Middle | 1546.32 | 1546.92 | 1588.09 | 1588.73 |
| Long | 1563.86 | 1564.47 | 1606.61 | 1607.25 |

In the configuration shown in Fig. 5, we tested the following three scenarios as shown in Fig. 7. In scenario 1, the routing was set so that both the C band and L band signals passed only through a link connected by the C + L band WSSs. In scenario 2, the L band signal was transmitted through the C + L band link, while the C band signal first passed through the C + L band link and then experienced the C band only link connected by the drop side C + L band WSS (upper right in Fig. 5) and C band WSS (lower left in Fig. 5). In scenario 3, the C band signal first passed through the C band only link, and then through the C + L band link. The C band signals experienced two spans, while the L band signals were transmitted for only one span due to limitations of the experimental equipment. In C + L band link, a C band 1-Tbps super-channel and an L band 1-Tbps super-channel are simultaneously propagated. Figure 7 shows network models corresponding to the three scenarios to help in understanding the differences among the scenarios.

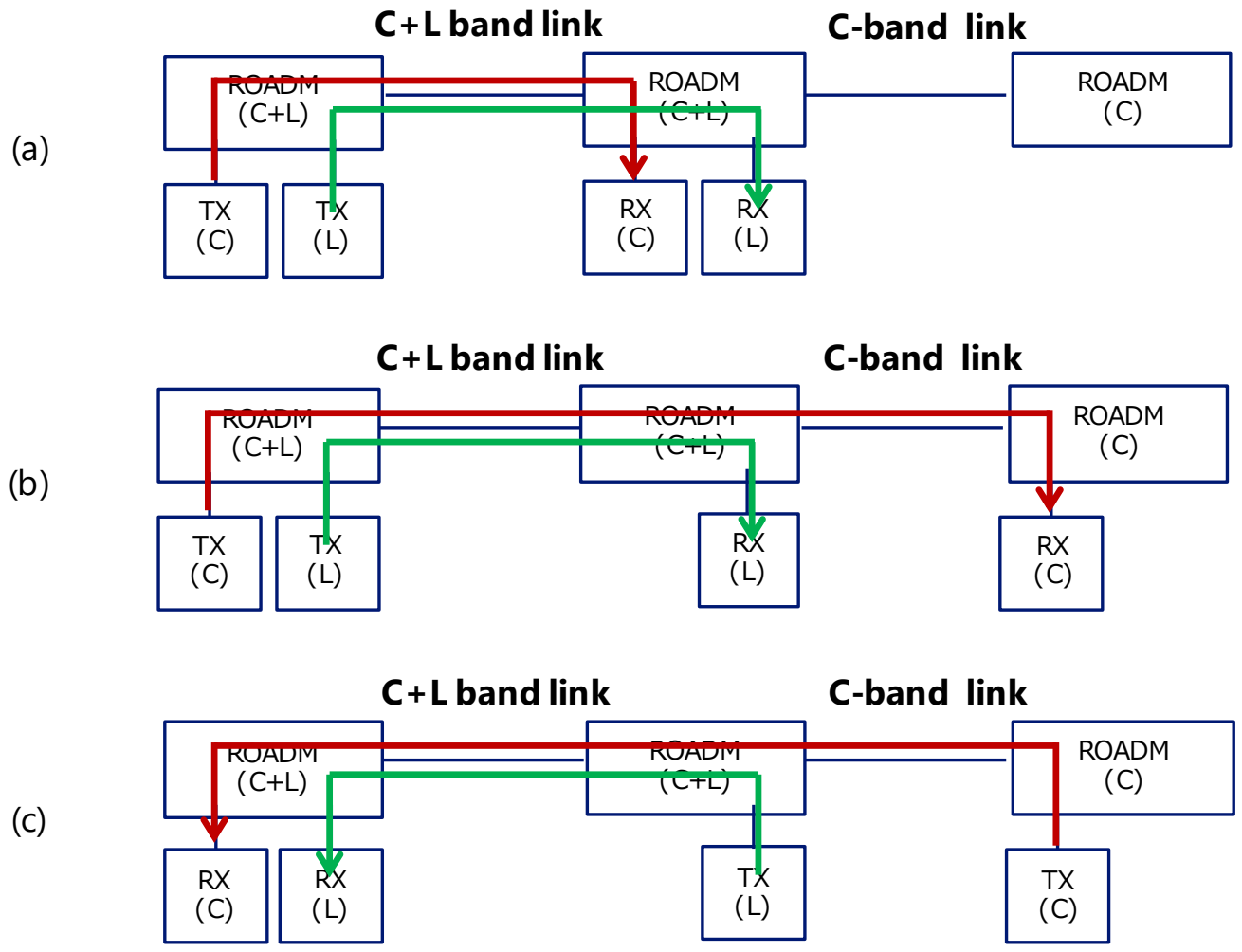

Fig. 7. Network models corresponding to (a) scenario 1, (b) scenario 2, and (c) scenario 3.

Figures 8(a) and (b) show the Q-factor margins from the forward error correction (FEC) threshold of the real-time coherent transceiver for C band signals and L band signals, respectively. The left plot is the result of a loopback configuration in which the Q-factor margins are measured by directly connecting the signal output to the signal input of the coherent transceiver as a reference. The other plots correspond to the above scenarios. The blue, green, and orange plots correspond to the results for each subchannel around the shorter edge, near the center, and around the longer edge of C and L bands, respectively. The insets show the typical constellations of the C band signal in Fig. 8. As shown in Fig. 8, no significant difference in the Q-factor margin was observed among scenarios 1, 2, and 3 for L band signals, whose Q-factor degradation was approximately 1.5 dB. This is because the signals are transmitted through only the C + L band link (the upper path in Fig. 5) for all scenarios. For C band signals, we could observe a 1.5-dB degradation of the Q-factor in scenario 1, where the signal was transmitted through only the C + L band link. This condition corresponds to one-span transmission. Further degradations of approximately 1 dB were observed in scenarios 2 and 3, which correspond to two-span transmission. For all cases in all scenarios, we confirmed that the post-FEC bit error rate (BER) is error free for each dual-carrier 1-Tbps super-channel. It should be noted here that the experiment was performed without any in-line amplifiers inside of the node, which supports the idea that we can omit the in-line amplifiers to simplify the node configuration.

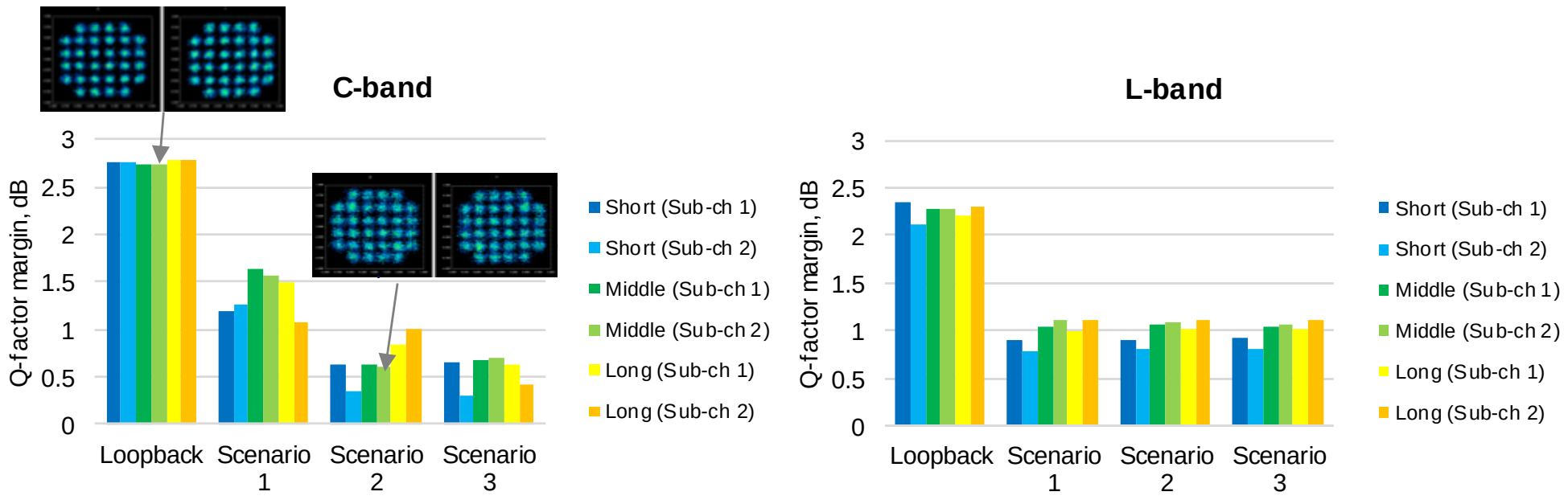

Fig. 8. Q-factor margin from FEC threshold of transponder in the three scenarios for C and L band signals.

We also conducted and additional experiment to support the elimination of amplifiers in the node. We varied the node input power at point A in Fig. 5 from 0 to 5 dBm/subchannel and measured the Q-factor margin for subchannel 2 of the dual-carrier 1-Tbps super-channel at near the center of the C band. Figure 9 shows the dependence of the Q-factor on the power at point A. We observed less than 0.2 dB of Q-factor change, which means that even if we installed an in-line amplifier with 5-dB gain between the WSS and MCS, we could only earn less than a 0.2-dB improvement.

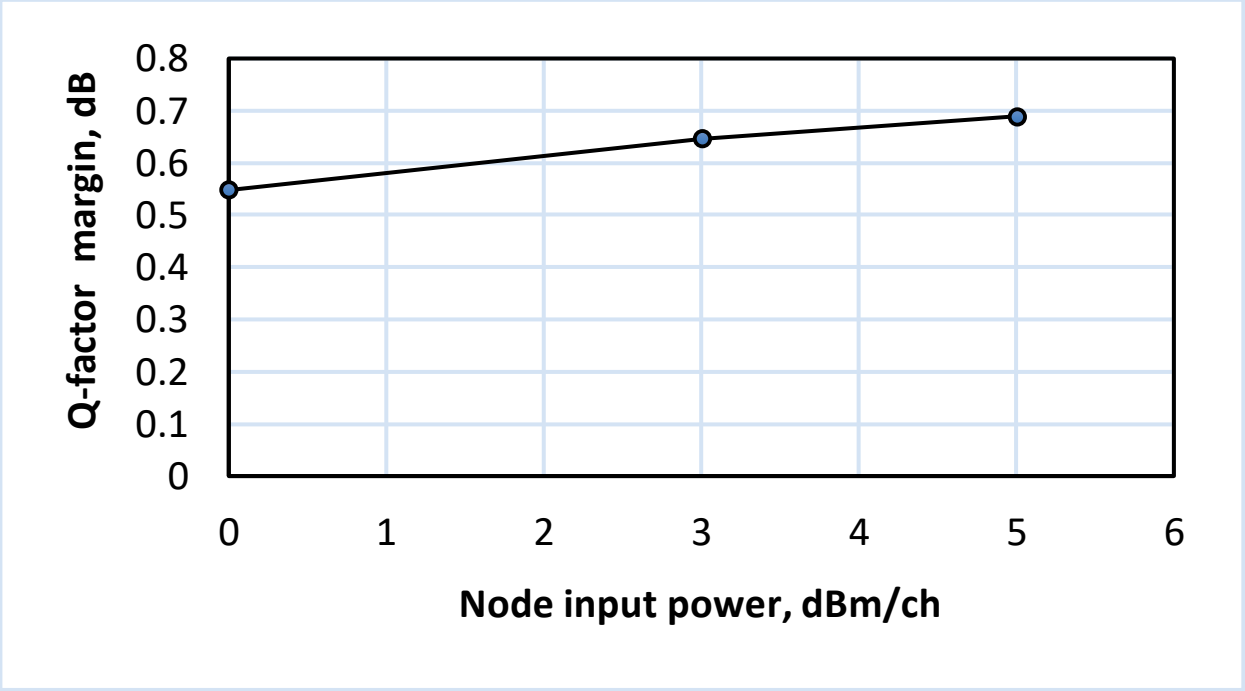

Fig. 9. Dependence of after-transmission Q-factor on node input power.

## 4. Conclusion

In this paper, we proposed a CDC-ROADM system that handles C + L band signals, which will be required in the high-baud-rate era, demonstrated its operation using real-time coherent transceivers, and verified satisfactory transmission with 1-Tbps dual-carrier 1 Tbps super channels. In C + L band CDC-ROADM, multiband switching devices are the key components, and we showed that the node construction can be simplified by using our C + L band WSSs and MCSs. In particular, the multiband MCS is attractive from an operational perspective because it operates as a band aggregator as well as a transponder aggregator. Furthermore, to simplify the node configuration, we clarified that an 8-client-port MCS can maintain the add/drop ratio at the same level as in the conventional C band CDC-ROADM system, where the optical amplifier for loss compensation in a node can be removed. As shown in this paper, we proved that the C + L band CDC-ROADM provides an essential network for the future high-baud-rate era.